# Enhanced Superconductivity in $Sr_2CuO_{4-y}$

T.H. Geballe
Dept of Applied Physics and Materials Science, Stanford University,
Stanford 94305, USA

M. Marezio
CRETA/CNRS, 38042 Grenoble cedex 9, France

A critical review of previous investigations of the superconductivity with enhanced $T_c \sim 95K$ found in $Sr_2CuO_{4-y}$ shows that new physics occurs in a highly overdoped region of the cuprate phase diagram. Moreover, evidence is adduced from the literature that 30% of the oxygen sites in the $CuO_2$ layers are vacant, a conclusion which is at odds with the universally made assumption that superconductivity originates in stoichiometric $CuO_2$ layers. While further research is needed in order to identify the pairing mechanism(s) responsible for the enhanced $T_c$, we suggest possible candidates.

## General Considerations

Liu et al. [1] and Yang et al. [2] have reopened the intriguing question concerning the origin of superconductivity in the cuprate superconductor $Sr_2CuO_{3+d}$. This material had been synthesized at very high pressures in the presence of a strong oxidizing agent by several groups starting with the initial discovery of superconductivity by Hiroi et. al. [3], and followed by a series of careful experiments initiated by Han et al[4], and Laffez et al [5]. But those measurements were soon suspect due to the possibility of contamination by the oxidizing agent employed. As noted below the method employed [1] made three major improvements in the synthesis : 1) alleged impurity contamination was disproved, 2) the need for long range oxygen diffusion was removed, and 3) the synthesis provided a good estimate of oxygen concentration.

Although the material is conventionally designated $Sr_2CuO_{3+d}$ we prefer to use the equivalent formula $Sr_2CuO_{4-y}$ because these superconductors have the same $K_2NiF_4$ structure as the $La_2CuO_4$ ["214"] cuprate family in which high temperature superconductivity was first discovered by Bednorz and Mueller over 20 years ago. Under the usual valence counting assumptions that Sr is +2, La is +3 and O is –2, the formal valence of the Cu cations in $Sr_2CuO_{4-v}$ is $+(2+p)$ where $p=2(1-v)$ is referred to as the concentration of doped holes and v as the concentration of oxygen vacancies. Annealing experiments following the synthesis have led to increases in Tc and to the identification of different periodic superstructures presumably related to further oxygen-vacancy ordering. The oxygen-vacancy concentration is variable. Liu et al [1] found that some superconductivity occurs over the range $v \approx 0.9$-$0.4$ with the maximum $T_c = 95K$ occurring at $v \approx 0.6$. As shown schematically in Fig 1 this opens an uninvestigated region in the phase diagram of high $T_c$ cuprate superconductivity.

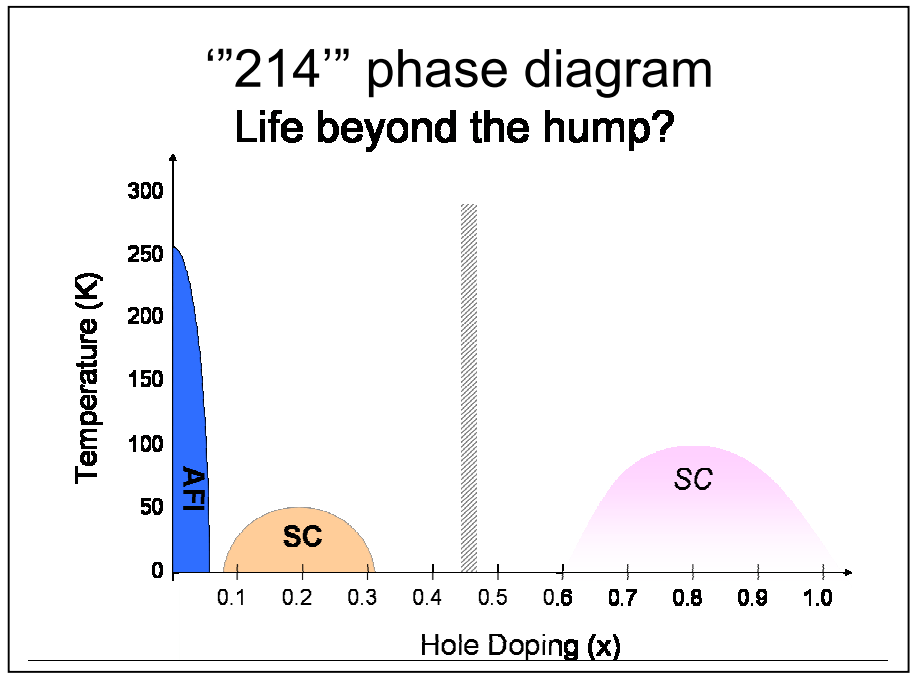

**Fig. 1:** Schematic representation of phase diagram of single layer cuprates that have been heavily doped. a = 3.756, c= 12.521. Han, et al (1994)

The doping mechanism is surprisingly effective in that the resulting $T_c$s are more than double those obtained in $(RE)_2CuO_4$ structures (RE being La or other rare earth elements) that are optimally doped using either cation substitutions or interstitial oxygen as hole doping agents. A significant difference in the superconducting behavior can be anticipated for $Sr_2CuO_{4-v}$ simply by noting p~0.16 in optimally doped $La_{1.84}Sr_{.16}CuO_4$ with $T_c \sim 40K$ and $T_c$ vanishes for p > 0.25, whereas p~0.8 in optimally doped $Sr_2CuO_{3.4}$ with $T_c = 95K$ [1]. Note, however, that in estimating the formal valence of Cu we assumed there is no peroxide remaining in the sample. This is a critical assumption because any non-recognized oxygen with -1 charge would lead to an overestimate of the Cu valence.[1]

This very high valence state of Cu alone portends large differences in the occupancy band filling. With 25% vacant oxygen sites in $Sr_2CuO_4$ there is still one electron per formula unit missing from the pd-bands. According to Harrison's early tight-binding bands for $CuO_2$ planes [8] this electron comes from the highest eg band, as in stoichiometric $La_2CuO_4$, providing a hole Fermi surface centered at the Brillouin Zone corner and no Fermi surface in the other bands. There were no unusual features in the band structure, which is not surprising considering that no strong local electron correlations were included.

**Unresolved Issues**

The location of the oxygen vacancies within the unit cell has yet to be firmly established. We offer reasons for believing that they are in the $CuO_2$ layers. This heretical idea is at odds with the universally accepted assumption that fully-stoichiometric $CuO_2$ layers are essential for obtaining the high $T_c$s. However, no theoretical or experimental work other than that discussed below has investigated the overdoped range well beyond p=0.35 where nonsuperconducting Fermi liquid behavior has been found.

Three important questions we address in this comment are: 1) What is the identity of the superconducting phase? 2) Where are the oxygen vacancies located? 3) Why is $T_c$ so much

---

[1] This assumption is chemically reasonable. Oxidation potentials show that Au, Ag and presumably Cu are oxidized to the trivalent state by peroxide. The $Cu^{+3}$ state obtained by heat treatment in oxygen is well known [6]. DeBacker et al [7] using iodometric and voltametric methods find that ~98% of the Cu in $NaCuO_{2-x}$ is trivalent, and furthermore virtually no remaining peroxide could be detected in photometry,

enhanced over that of all other "optimally" doped 214 cuprates?    Because presently available samples are all multiphase it is not possible to answer these questions with certainty. However, on the basis of an analysis of all the existing data, we tentatively conclude that: 1) the same majority phase which is present for all the various synthetic protocols is (contrary to what has been previously asserted) the superconducting phase; 2) the structural evidence that the O vacancies are primarily in the Cu-O layer in the majority phase thus implies that high temperature superconductivity arises in planes with very different structure than in more familiar materials. The authors of references [1] and [2], and the original discoverers of the material, Hiroi et al. [3], assumed that the oxygen vacancies were located in the rock salt blocks, that is in the $(SrO)_2$ double layers. This assumption, although unquestionably reasonable, is not based upon any direct experimental evidence; in fact we find that the evidence suggests otherwise. Concerning question 3, we speculate briefly on various roles the in-plane O vacancies may be playing in the mechanism of superconducting pairing.

**Identifying the superconducting phase in multiphase samples**

Minority phases have, indeed, been found for all methods of synthesis including that employed by Hiroi et al. [3], Liu et al. [1] and by Han et al. [4]. The minority-phase-explanation gained credence a few years later from experiments by Scott et al. [9]. Those authors used a high-pressure synthesis method, claimed to be similar to that described in references [3] and [4]. However, they used $KClO_3$ for oxidizing the starting $Sr_2CuO_3$ material while Hiroi et al. [3] and Han et al. [4] had used a much stronger oxidizing agent, $KClO_4$. With a scanning SQUID microscope, Scott et al. [9] identified a minority (~3%) superconducting oxychloride cuprate phase in their macroscopically inhomogeneous sample that was superconducting and could account for the small zero-field-cooled superconducting signals that were observed. They further suggested that the same minority phase could have been responsible for the superconductivity observed in the previous work by Hiroi et al. [3], Han et al. [4], and others. Although a direct comparison is not possible, the magnitude of the Scott et al. [9] signals are at least a factor of 5 smaller than the ones reported by Han et al. [4] in Fig 2, and the lattice constants obtained from powder x-ray diffraction are markedly different. Nevertheless after their suggestion there was a cessation of further research until Liu et al [1] employed $SrO_2$ as the oxidizing agent in their high pressure synthesis. This elegant procedure eliminated the possibility of oxychloride contamination. Furthermore, since fine-grained mixtures of the constituents were employed, the incomplete diffusion encountered by Scott et al. was avoided and the authors were able to estimate the oxygen content of their sample from the initial composition.

In Fig 2 we compare the superconducting behavior of the Liu et al. [1] samples with that of Han et al. [3] samples. It should be noted that the samples being compared here were made more than a decade apart in time in different laboratories using different oxidizing agents. If the superconductivity properties were due to an unidentified minority phase because of Scott's suggestion, it is surprising that remarkably similar signals were found in the two investigations. In both cases the $Sr_2CuO_{4-v}$ phases were synthesized under high pressures ~6GPa and T≈1100 C (Liu) and 900 C (Han) and retained as metastable phases at room temperature and atmospheric pressure. It is not surprising that the quite different methods of synthesis resulted in the formation of different minority phases and in different degrees of stability [the Han samples were retained at higher temperatures before reverting to the low-pressure phase $Sr_2CuO_3$ as shown in Fig 2.

It is however significant that the same majority phase with almost identical Tcs was found and that the subcell lattice constants were reasonably close. Both sets of samples contained incommensurate periodic superstructures. The periodicity changed upon the subsequent annealing at intermediate temperatures that resulted in the higher $T_c$s shown in Fig 2 and then reverted to the starting non-superconducting phase after annealing at still higher temperatures. In the analysis of the Han samples [Han et al. [4], Wang et al. [10], and Zhang et al. [11] ] 3 different periodic structures were found during the annealing, along with traces of impurity phases ($SrO_2$, SrO, and CuO), whereas Liu et al. [1] reported 4 incommensurate periodic superstructure phases described below. All these phases contain high concentrations of oxygen vacancies but it is only in the majority phase that it has been established that they are located in the $CuO_2$ layers in their samples [12].

Field-Cooled Meissner

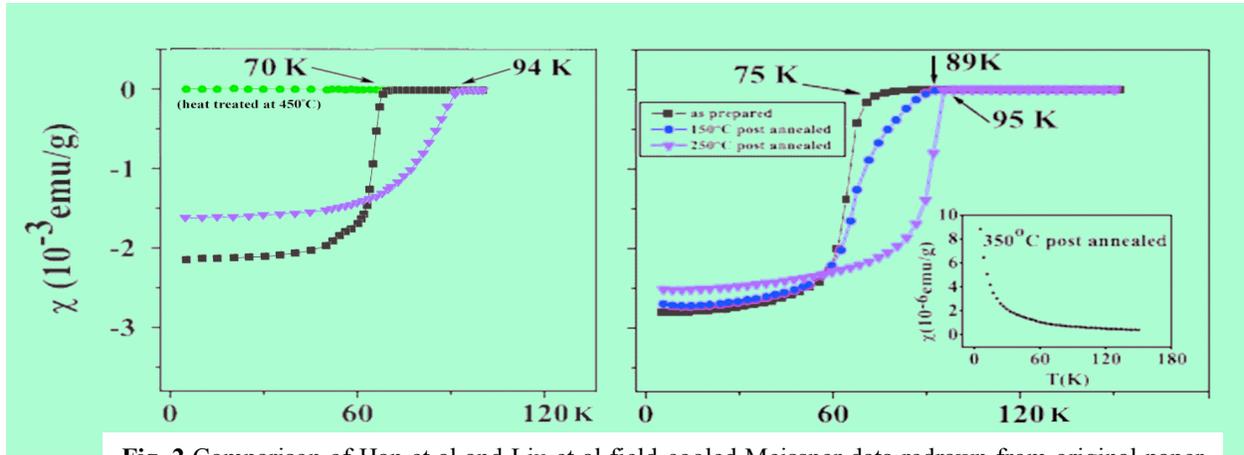

**Fig. 2** Comparison of Han et al and Liu et al field-cooled Meissner data redrawn from original paper to be on same scale. The 450C anneal of Han data shown in green retained the majority phase and was not super-conducting. The 350° anneal of the Liu data (shown in insert with different scale) also retained the majority phase and was not superconducting. a = 3.795, c = 12.507. Q.Q. Liu et al (2004)

Both the Han et al. [4] and Liu et al. [1] samples were characterized by x-ray and electron-diffraction and imaging techniques. The Han et al. [4] samples were used by Wang et al. [10] and by Zhang et al. [11] in further research. Also, the Liu et al. [1] and Yang et al. [2] research was done on the same set of samples. We will use "Han" and "Liu" at times to cover the respective properties of the two sets of samples. The majority phase found by both groups in the as-prepared superconducting samples was tetragonal I4/mmm with lattice constants a = 3.756 Å, c = 12.521 Å (Han) and a = 3.795 Å, c = 12.507 Å.[2]   (Liu). Liu el al. [1] and Yang et al. [2] noted no change in the majority phase upon annealing to 250 C which resulted in a rise in Tc from 75K to 95K, as shown in Fig 2, but they did observe significant changes in the minority phases upon annealing. They therefore attributed the different T cs found to the different the minority phases that appeared. On the other hand, as outlined below, Han et al. [4] found subtle changes in the majority phase upon annealing first at 310 C, and subsequently more evidence upon annealing to 450 C that indicated differences in oxygen ordering in the majority phase and could account for the different $T_c$s.

---

[2] The powder x-ray diffraction studies revealed that the as-prepared phase, yielded by the hihgpressure synthesis, had the tetragonal K2NiF4 structure with space group symmetry I4/mmm with ap=bp and cp lattice parameters. The electron diffraction studies indicated that the crystal symmetry was actually orthorhombic with space group symmetry Fmmm and a≈ b ≈ 5√2$a_p$ and c ≈ $c_p$.

In spite of the remarkable similarity of the $T_c$ behavior, Liu et al. [1] dismissed the Han et al. [4] results as being multiphase with a very small superconducting fraction. That conclusion is supported neither by a comparison of the $T_c$s shown in Fig 2 nor by the similar magnitudes of the Meissner (flux expulsion) signals observed in the field-cooled measurements. The field-cooled susceptibilities (Fig 2) are measured in low fields of 10 Oe and 20 Oe for the Han et al. [4] and Liu et al. [1] samples, respectively. We believe that the same phase is responsible for the superconductivity in both cases, although more work is needed before this can be established with certainty.

Wang et al. [10] and Zhang et al. [11] used the Han et al. [4] samples to study the evolution of the modulation period in the majority phase. Transmission electron microscopy and diffraction studies (see Fig 4 of [10]) show that the discommensuration period changes (in √2 units of the I/4mmm unit cell) from 4.8 ± 0.05 for the $T_c$=70K sample to 5.2±0.05 for the $T_c$=94K sample. With further annealing to 450 C that results in the loss of superconductivity, the periodicity becomes 5.0±0.1 indicating that it has become commensurate. As noted below there is further evidence in the reduced density of states that is found in the 1s absorbtion edge of the oxygen ions.

Liu et al [1], by examining a large number of grains of their as-prepared samples using electron diffraction, found that 20% were monoclinic, space group C2/m, with crystallographic axis a ≈ 5√2$a_p$, b ≈ $c_p$, c ≈ √26√2/2$a_p$ and _ = 101.3°, the index p referring to the I4/mmm subcell.

The remaining (majority) 80% of the grains had the Fmmm structure with a ≈ b ≈ 5√2$a_p$, and c ≈ $c_p$. Upon subsequent annealing at 150 °C most of the monoclinic phase transformed to an orthorhombic phase Cmmm with unit cell parameters a ≈ $c_p$, and b ≈ c ≈ 5√2$a_p$. This implies that at that transition the _ angle changes from 101.3° to 90°. After the 250 °C annealing, all grains with the C2/m and Cmmm structures changed to a new orthorhombic structure belonging to space group symmetry Pmmm with b ≈ c ≈ 4√2$a_p$, a ≈ $c_p$ to which the 95K superconductivity was attributed. We would like to point out that the monoclinic structure C2/m found by Liu et al still has the K2NiF4 structure, but with a _ angle equal to 101.3 The rotation of the crystallographic axes on going from C2/m to Cmmm, to Pmmm is due to the choice of conventional space groups.

As already noted, Liu et al. [1] assume that different $T_c$s found upon annealing are due to the different minority phases that are formed and further that the vacancies must be in the apical oxygen sites in the $(SrO)_2$ layers in order to maintain the stoichiometry of the $CuO_2$ layers they assume to be an essential requirement for obtaining high $T_c$s. The experimental evidence in our opinion indicates that both these assumptions are incorrect.

First, let us consider the magnitude of the Meissner field-cooled experiments. Usually, flux trapping in multiphase material prevents full expulsion of the field and furthermore a distribution of pinning centers is not conducive to sharp transitions. In a quantitative investigation of flux expulsion in a related single phase cuprate, $Sr_{.9}La_{.1}CuO_2$ Kim et al. [13] found the field-cooled Meissner signal to be roughly 30% of that expected for perfect diamagnetism. Even if that same 30% ratio were applicable to the minority phases found [1, 2], which is unlikely because the ED images show defects such as twin and grain boundaries, the rather sharp signals in Fig 2 would be expected to originate from only a few volume percent of the sample rather than the 15% that is reported .

Secondly it is highly unlikely that Han et al [4] would have failed to detect the monoclinic phase had it constituted 20% of their sample. It is more likely that the different syntheses used by

Han and by Liu yielded different minority phases.[3] The most reasonable explanation for the similarity of the data in Fig 2 is that the majority Fmmm phase, common to both the Han and Liu samples, is superconducting. Further support for attributing the superconductivity to the Fmmm phase is given below.

An alternative to high pressure synthesis is the use of epitaxy to deposit single phase films with the desired structure. Pioneering work has been done using molecular beam epitaxy to deposit the insulating $Sr_2CuO_3$ orthorhombic films and subsequently converting them with ozone oxidation to the tetragonal $K_2NiF_4$ structure [14]. However, the relevance to the present discussion is not clear since the films had significantly larger c-axes ≈ 13.55 Å than the bulk samples being considered here. As pointed out by H. Yamamoto [15] a c-axis longer than that of $La_2CuO_4$ (c ≈ 13.3 Å) is to be expected for a simple substitution of Sr for La. However $Cu^{+2.8}$ has a smaller radius than $Cu^{+2}$ which can explain the reduction of the c-axis. It should also be noted that the best film had a smeared out transition with an onset $T_c$ of ~20K as measured by the magnetic susceptibility.

**Location of oxygen vacancies**

Shimakawa et al. [12] used powder neutron diffraction data on a sample provided by Hiroi et al [3] and a simplified model of the periodic modulation to carry out the structural refinements of the majority phase. Those refinements showed that the oxygen vacancies are essentially all in the $CuO_2$ layers. The SrO layers are nearly flat precluding the vacancies being in them. The sample used had appreciable but somewhat poorer superconducting behavior than those shown in Fig 2. The onset of $T_c$ was 65 K and was relatively sharp; 2/3 of the Meissner signal of 5.6% occurred by 60 K. Following the neutron work the sample was powdered and the volume fraction determined by ac susceptibility was 7.6%. The poorer signals might have resulted from some degradation due to the elapsed time between preparation and measurement that is sufficient to rule out the suggestion of Scott [9] whose samples were reactive when exposed to air. Although it would be helpful to have measurements on a better superconducting sample the refinement of the neutron data indicates very strongly that the vacancies are on the $CuO_2$ layers.

---

[3] If further high pressure synthesis is unable to produce single phase material it should still be possible to make a conclusive assignment. Heat capacity measurements would be very helpful but none have been reported. A simple [at least in principle] way of making a definite identification would be to separate the superconducting grains from the rest by exposing a powdered sample cooled below $T_c$ to a field gradient.

The simulation of the electron diffraction data by Zhang et al. [11] is consistent with the neutron data. The two different Cu-Cu distances found in the copper oxygen plane that can only be due to the presence and absence of oxygen vacancies between the copper ions.

**Evidence for competitive ordering**

Electron energy loss spectroscopy [EELS] was used to measure the oxygen 1s spectrum in both the Han and the Liu samples. The magnitude of the low energy pre-peak near 530 eV is a measure of the local density of states and thus is sensitive to the conduction electron density. Both sets of data show qualitatively a similar pre-edge peak for the as prepared Fmmm majority phase. Yang et al. [2] using the Liu samples found no perceptible change in the low energy peak of the Fmmm majority phase when measured in their as-prepared sample and upon post annealing to reach 95K, reinforcing their conclusion that the majority phase was not superconducting. Upon annealing to 350 C where superconductivity disappears they found that 30% of the sample had converted to the stable $Sr_2CuO_3$ phase and found as expected a marked decrease in the magnitude of the pre-edge peak of that insulating phase. No data are presented for the 70% of the remaining Fmmm phase. This is unfortunate because Wang et al.'s results show that when the $T_c$ disappears in the Han-prepared samples upon annealing at 450 C the Fmmm phase is present with a markedly reduced pre-edge peak. This is evidence for a much reduced density of states. Furthermore the periodicity modulation becomes commensurate within experimental error as already mentioned above. The Wang et al. [10] data thus suggest that the disappearance of superconductivity in the Fmmm phase is due to the formation of a charge density wave, possibly a metal insulator transition. If our interpretation of Wang et al's data is correct Yang et al. [2] should have found similar results had they examined their Fmmm phase after the loss of superconductivity upon annealing at 350 C anneal.

**Possible mechanisms**

The arguments given above lead to the conclusion that the $CuO_2$ layers contain oxygen sites that are ~30% vacant and that a challenging new region of the phase diagram exists with $T_c$s twice that of the well investigated 214 cuprates. It is beyond the scope of this article to present theoretical models but below we can suggest two scenarios that are at least conceivable.

1) Negative-U centers

It well established that some local defects are ``negative U'' centers -i.e. sites for which there are two preferred charge states differing by a pair of electrons. If, in addition, the energy scale for quantum fluctuations between these two states is sufficiently large, they can be a source of superconducting pairing [16-18]. Such a negative-U mechanism of superconductivity has recently been found to account for the anomalously high $T_c$ in Tl-doped PbTe, where the Tl acts as a ``valence skipping atom, fluctuating between valence $Tl^{+1}$ and $Tl^{+3}$ [19]. Furthermore, it has been suggested that Tl acting in a similar manner in the cuprate Tl2201 is responsible for the enhanced superconductivity found in that compound [20].

In view of the above we are led to the speculative idea that the oxygen vacancies themselves might be negative-U centers in $Sr_2CuO_{(4-v)}$. Oxygen vacancies are known to bind electron pairs in the binary alkaline earth oxides such as BaO. These pairs exist as diamagnetic localized states below the chemical potential and are known as F-color centers. Because they are tightly bound, they are ineffective in producing superconductivity. On the other hand, cluster calculations in $SrTiO_3$ [21-23] show that in this ternary perovskite there is a delicate balance between the electron pair's being localized around an oxygen vacancy and being delocalized throughout the cluster. There is even circumstantial evidence that such O vacancies may be essential for superconductivity found in SrTiO$_3$ [24]. If the O vacancies in Sr$_2$CuO$_{(4-v)}$ behave like the O vacancies in SrTiO$_3$, they could play a significant role in the superconducting pairing. The simulation of the electron diffraction data by Zhang et al [11] as mentioned above gives alternating short and long Cu–Cu bonds suggesting alternating oxygen vacancies. Overlap in such a case would result in transport by paired electrons. Model calculations show that a narrow band of paired states will become superconducting [25].

2) Optimum inhomogeneity.

Another possibility is that the oxygen vacancy clustering might provide a form of ``optimal inhomogeneity" for superconductivity. The evidence for competitive charge-density-wave formation discussed above suggests the possibility of the oxygen vacancies being organized to provide micro metalinsulating regions. There are as yet untested theoretical arguments that Tc can be enhanced under suitable circumstances in a system consisting of a composite where regions with strong pairing but small coherence scale proximity coupled to more metallic regions with little or no pairing but strong electron itineracy [26]. From this viewpoint, it may not be the average Cu valence, but the Cu valences in distinct regions of the Cu-O planes that is relevant to the mechanism of the high temperature superconductivity.

**Conclusions**

A metastable superconductor with non-stoichiometric CuO$_2$ layers exists in an unexplored overdoped region of the phase diagram of cuprate superconductors. The enhanced T$_c$ of 95 K found in Sr$_2$CuO$_{4-v}$ for v ≈ 0.6 that corresponds to an average copper valence of ~ 2.8+. An important new synthesis introduced by Liu et al. [1] removes the possibility that the observed superconductivity is due an oxychloride cuprate contaminant as had been previously suggested. There is some uncertainty in interpreting the data due to the fact the samples are multiphase. However comparison of remarkably similar superconducting behavior of samples made over a decade apart in different laboratories by different methods demonstrates a robustness that identifies the majority Fmmm phase as being responsible for the superconductivity. There is evidence for the appearance of a competitive CDW ordered state in the Fmmm phase when it is annealed at a high enough temperature to cause the disappearance of superconductivity. New mechanisms are needed to account for the enhanced superconductivity. Two possibilities are suggested. One involves a band of paired electrons formed by the overlap of negative-U oxygen vacancies. The second involves pairing in the $CuO_2$ layer that is enhanced either by an optimal inhomogeneous distribution of oxygen sites or by a distribution of vacancy rich/vacancy poor regions.

Acknowledgements: We are grateful to Prof C. Q. Jin for sharing his results and for interesting discussions. We have greatly benefited from discussions and comments from many colleagues and would particularly like to acknowledge helpful comments from Steve Kivelson, Erez Berg, Ivan Bozovic, Doug Scalapino and Art Sleight. We thank Walter Harrison for his tight binding calculation and would like to acknowledge helpful information provided by J. F. Mitchell concerning the neutron studies. One of us (THG) would like to acknowledge support from the Air Force Office of Scientific Research.